\documentclass[a4paper,11pt]{article}
\usepackage{pos}

\usepackage{natbib}
\title{Study of the upgraded EUSO-TA performance via simulations}
\ShortTitle{Study of the upgraded EUSO-TA performance via simulations}

\author*[a]{Francesca Bisconti}

\affiliation[a]{INFN, Sezione di Roma Tor Vergata \\ Roma, Italy}

\onbehalf{for the JEM-EUSO Collaboration\\[-1mm]{\normalsize \normalfont (a complete list of authors can be found at the end of the proceedings)}}

\emailAdd{francesca.bisconti@roma2.infn.it}

\abstract{The EUSO-TA ground-based fluorescence detector of the JEM-EUSO program, which operates at the Telescope Array (TA) site in Utah (USA), is being upgraded. In the previous data acquisition campaigns, it detected the first nine ultra-high energy cosmic ray events with the external trigger provided by the Black Rock Mesa fluorescence detectors of the Telescope Array (TA-BRM-FDs). Among the upgrades, there is the installation of a trigger algorithm for the independent detection of cosmic ray air showers and upgraded electronics.
A simulation study was developed to understand the performance of EUSO-TA in the new setup and different background conditions. This study allowed us to estimate the detection limit of the ground-based detector, which can be used to extrapolate the detection limit for a balloon-based detector. Moreover, it provided estimations of the expected trigger rates for ultra-high energy cosmic rays.
In this work, the description of the simulation setup, the method developed to rescale the energy of the cosmic rays to account for the portion of air shower actually observed rather than the whole one, and the results in terms of detection limit and trigger rates, are reported.}

\ConferenceLogo{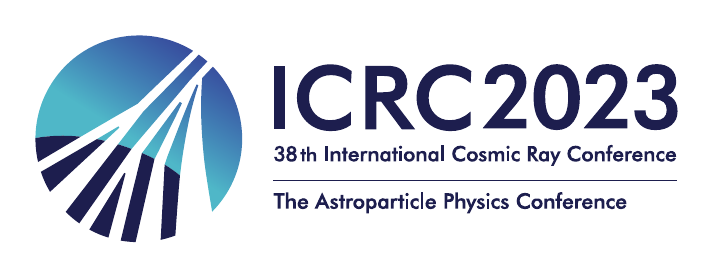}

\FullConference{%
38th International Cosmic Ray Conference (ICRC2023)\\
  26 July - 3 August, 2023\\
  Nagoya, Japan}

\usepackage{siunitx}
\usepackage{lineno}

\usepackage{xspace}
\newcommand{\Offline}{\mbox{$\overline{\textrm%
			{Off}}$\hspace{.05em}\protect\raisebox{.4ex}%
		{$\protect\underline{\textrm{line}}$}}\xspace}

\begin{document}
\maketitle

\section{The EUSO-TA detector and observations of UHECRs}\label{sec:euso-ta_detector}

EUSO-TA \cite{bib:euso-ta} is an experiment of the JEM-EUSO program \cite{bib:jem-euso} developed and operated to validate the observation principle and the design of a particular kind of detector, capable of observing extensive air showers (hereafter ``showers'') induced by Ultra-High Energy Cosmic Rays (UHECRs) and laser pulses. UHECRs can be detected by observing, at nighttime, the UV fluorescence and Cherenkov light emitted when showers cross the atmosphere. EUSO-TA is installed at the Telescope Array (TA) \cite{bib:ta} site in Utah (USA), in front of the Black Rock Mesa Fluorescence Detectors (TA-BRM-FDs).

The EUSO-TA optics consists of two Fresnel lenses of 1~m diameter and 8~mm thickness, made of Poly(methyl methacrylate) - PMMA \cite{bib:lenses}. The focal surface consists of one Photo-Detector Module (PDM), composed of $6\times6$ Hamamatsu Multi-Anode Photo-Multiplier Tubes (MAPMTs, model R11265-M64) \cite{bib:mapmt} with $8\times8$~pixels of 2.88~mm side each. The field of view (FOV) of one pixel is  \ang{0.2}$\,\times\,$\ang{0.2}, making a total FOV of $\sim$\ang{10.6}$\,\times\,$\ang{10.6}. A 2~mm thick UV band-pass filter (in the range $290-430$~nm), is glued on top of each MAPMT. 

Data are sampled in $2.5\,\mu\mbox{s}$ windows (called Gate Time Units - GTUs). Prior to the ongoing upgrade, the readout was performed by one 64-channel SPACIROC1 ASIC chip \cite{bib:spaciroc1} per MAPMT, with a dead time at the beginning of each GTU of 200~ns and 30~ns double pulse resolution.
The elevation of EUSO-TA can be set from \ang{0} to \ang{30} above the horizon, whereas its azimuth is fixed at \ang{53} from North counterclockwise, pointing to the Central Laser Facility (CLF) of TA.

The location of EUSO-TA in front of the TA-BRM-FDs makes it possible to use the external trigger provided by the TA-BRM-FDs to record cosmic ray events. Four data acquisition sessions in 2015 ($\sim123$~hours) and one in 2016 (for to a total of $\sim140$~hours) were performed using the TA-BRM-FDs trigger. Nine showers were found in the EUSO-TA data collected in 2015, while several tens that crossed the FOV were not, as the EUSO-TA spatial and temporal resolutions were optimized for observations from space.

To improve the data acquisition, an upgrade of EUSO-TA is ongoing to allow for remote operations with self-trigger capability, defining the so-called EUSO-TA2 \cite{bib:euso-ta_icrc2023}. 
The electronics will be updated with SPACIROC3 boards~\cite{bib:spaciroc3}, with a dead time at the beginning of the GTU of 50~ns and 5~ns double pulse resolution. 
Moreover, the detector will have self-triggering capabilities (level-1 trigger logic) \cite{bib:trigger}. 
The data read-out will be possible on three timescales: 2.5~$\mu$s for the observation of showers; 320~$\mu$s to follow the evolution of fast atmospheric events; and 40.96~ms for slow events such as meteors and strange quark matter (strangelets). 

In this work, we describe the methods used to estimate the detection limits, in terms of energy and distance to shower, of EUSO-TA with the internal level-1 trigger.  
Moreover, the expected UHECR trigger rates are reported.

\section{Correction of the energy}\label{sec:energy_correction}

It is important to understand the EUSO-TA detection limit in terms of shower energy and the distance to the shower along the telescope optical axis. However, the relatively small FOV of EUSO-TA allows observation of only a portion of the shower that in most cases does not include the portion of the shower with the maximum number of particles, resulting in a smaller signal than would be observed if the shower maximum was in the FOV.
To account for this, we define a so-called equivalent energy relating the energy deposit of the partially-observed shower to the energy deposit at the shower maximum.

To compute the equivalent energies, we determine conversion factors to apply to the true (simulated) energies using a collection of shower simulations performed using CONEX \cite{bib:conex}. 
The simulated showers have several energies ($\mathrm{E=10^{17}-10^{20}}$~eV, with steps of $\mathrm{\log E(\mbox{eV})=0.1}$) and zenith angles ($\theta=\ang{0}-\ang{65}$, with steps of \ang{5}). Several elevation angles of EUSO-TA (\ang{10}, \ang{15}, \ang{20}, and \ang{25}) and distances from the detector ($\mathrm{D=1-50}$~km, in steps of 1~km) were considered. 
The energy conversion factor, $\mathrm{f_{eq}=(dE/dX)_{obs}/(dE/dX)_{max}}$, is defined as the ratio of the energy deposit per unit altitude of the shower at the observed point and that at the maximum. 
To reduce fluctuations in the results, 20~showers were simulated with each combination of energy and zenith angle, and for each one, the mean conversion factors were calculated for different distances and elevation angles of EUSO-TA. 
\begin{figure*}[h!]
	\centering
	\includegraphics[height=3.8cm]{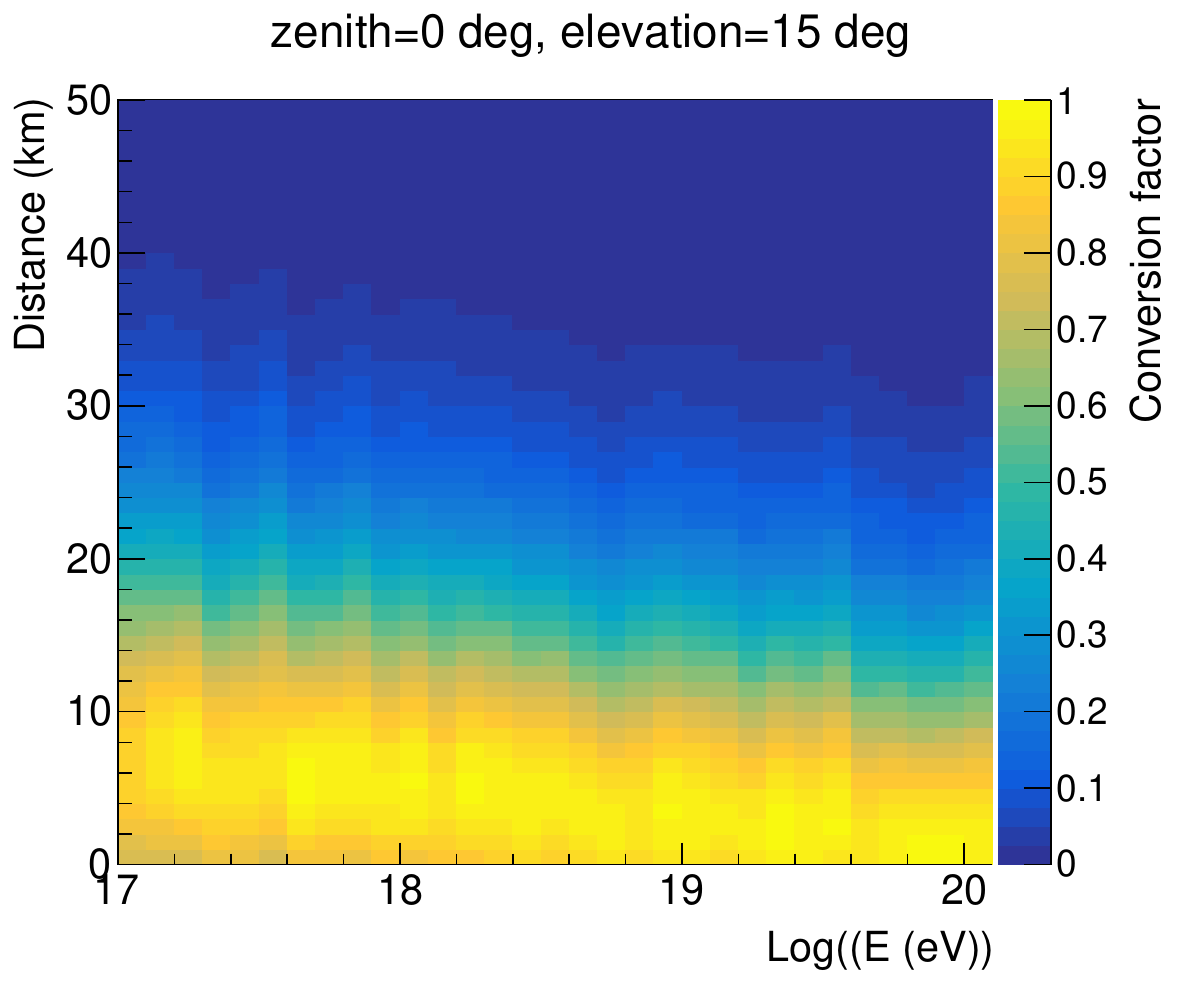}
	\includegraphics[height=3.8cm]{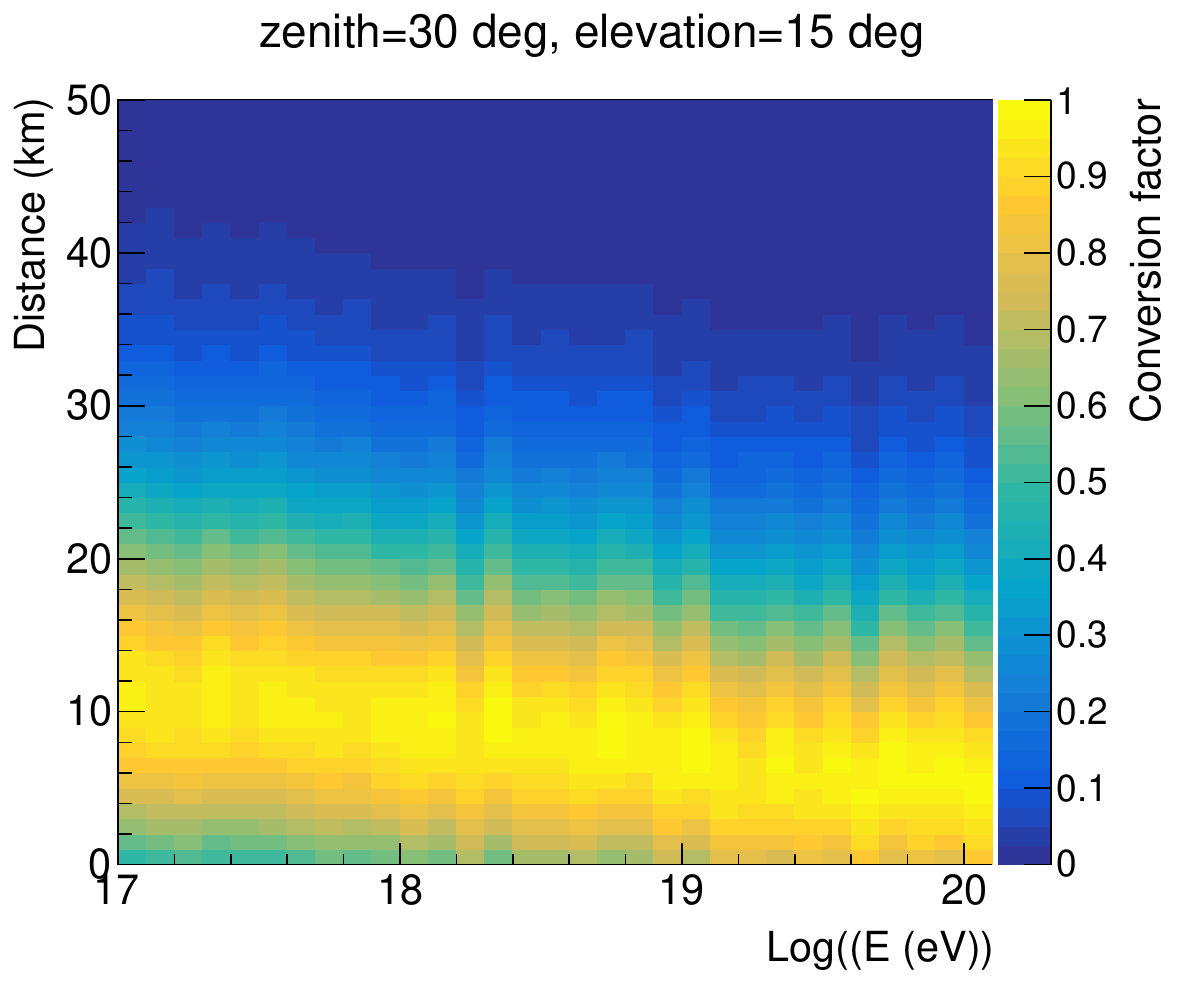}
	\includegraphics[height=3.8cm]{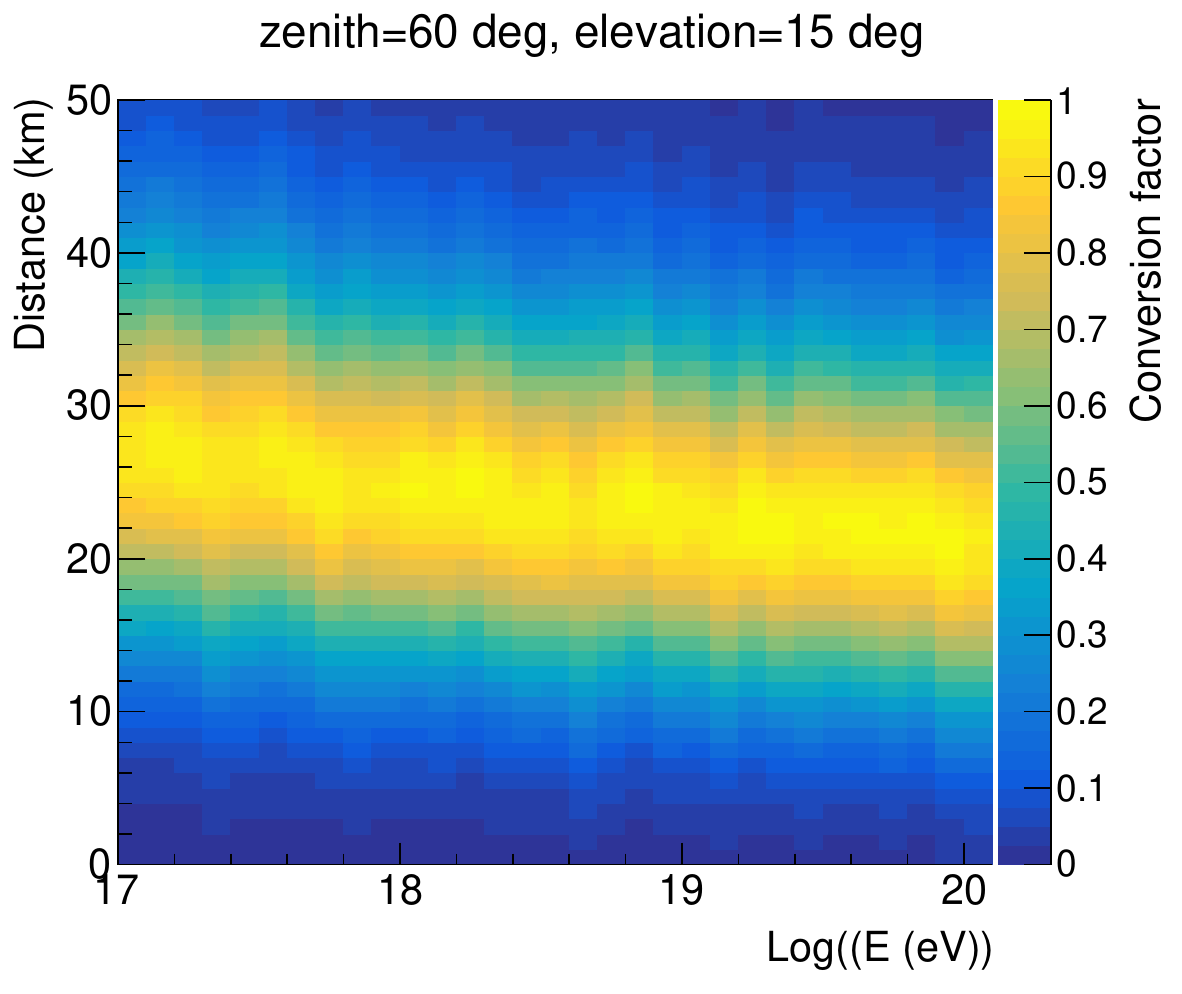}
	
	\caption{Energy conversion factors as a function of the shower energy and distance from the detector. The plots refer to \ang{15} elevation angle and \ang{0}, \ang{30}, and \ang{60} zenith angles.} \label{fig:factors}
\end{figure*}
Values of the energy conversion factors are displayed in Figure~\ref{fig:factors} as a function of the shower energy and distance from the detector. The plots refer to elevation angles of \ang{15}, and zenith angles of \ang{0}, \ang{30}, and \ang{60}. The conversion factors are near 1 in cases where the shower maximum is within the FOV, and become smaller as the maximum moves farther from the FOV. As expected, by increasing the energy and decreasing the zenith angle, the shower maximum is closer to the detector, i.e. at lower altitudes. 

To estimate the energy of the partially-observed shower, the atmospheric transmission has to be taken into account, as for a given elevation angle, the slant depth along the EUSO-TA optical axis that intersects the shower axis may be different from that pointing to the shower maximum. The Linsley parametrization used in the CORSIKA simulation software \cite{bib:corsika} was used to retrieve the atmospheric depths at given altitudes. For a given atmospheric depth $\mathrm{X}$, the atmospheric transmission is given by $\mathrm{T=\exp(-X/\Lambda)}$, 
where $\mathrm{\Lambda}$ is the mean free path for Rayleigh scattering, which, in the near-UV, is $\mathrm{\Lambda(350~nm)=1700~g/cm^2}$. 
With the atmospheric slant depth up to the observed point of the shower $\mathrm{X_{obs}^{slant}}$ and the one up to the shower maximum $\mathrm{X_{max}^{slant}}$, it is possible to calculate the corresponding atmospheric transmission $\mathrm{T_{obs}}$ and $\mathrm{T_{max}}$. 
The atmospheric transmission is corrected for the fact that the number of photons arriving at the detector is inversely proportional to the square of the distance.
The atmospheric correction factor $\mathrm{f_{atm}=T_{obs}/T_{max} \cdot D_{max}^2/D_{obs}^2}$ is the ratio between the atmospheric transmission to the observation point and the shower maximum, both corrected for the corresponding distances $\mathrm{D_{obs}}$ and $\mathrm{D_{max}}$. The conversion factor may be equal to 1 when the shower maximum is in the FOV, but in general, it is greater or less than 1, depending on the shower direction. The equivalent energy corrected for the atmospheric transmission becomes $\mathrm{E_{eq,atm}=f_{eq} \cdot f_{atm}\cdot E_{sim}}$.

%
\section{Simulation set}\label{sec:simu_set}
For this analysis, a set of 10000~shower simulations was performed with CONEX, using the QGSJETII-04 hadronic interaction model, protons as primary particles, zenith angles in the range $0^{\circ}-90^{\circ}$ drawn from the isotropic flux on a flat surface, i.e. $\mathrm{dN/d\cos(\theta) \sim \cos(\theta)}$ and random azimuth angles in the range $0^{\circ}-360^{\circ}$. The energy range was $1\times 10^{17}-1\times 10^{20}$~eV with spectral index $-1$ in logarithmic scale in order to have statistics also at the highest energies. The CONEX showers were then processed with the \Offline framework~\cite{bib:offline} to perform the production and propagation of fluorescence and Cherenkov photons from the shower to the detector, and to simulate the detector response with the level-1 trigger.
The showers were distributed over a large area of $36\times28\:\mbox{km}^2=1008\:\mbox{km}^2$ centered at half the distance between EUSO-TA and the CLF. The elevation angles chosen for EUSO-TA were \ang{10}, \ang{15}, \ang{20} and \ang{25}, while the simulated background levels were 1 and 1.5~counts/pixel/GTU.
\section{Detection limit}\label{sec:simu_limit}
\begin{figure*}[thp]
	\centering
	\includegraphics[width=6.cm]{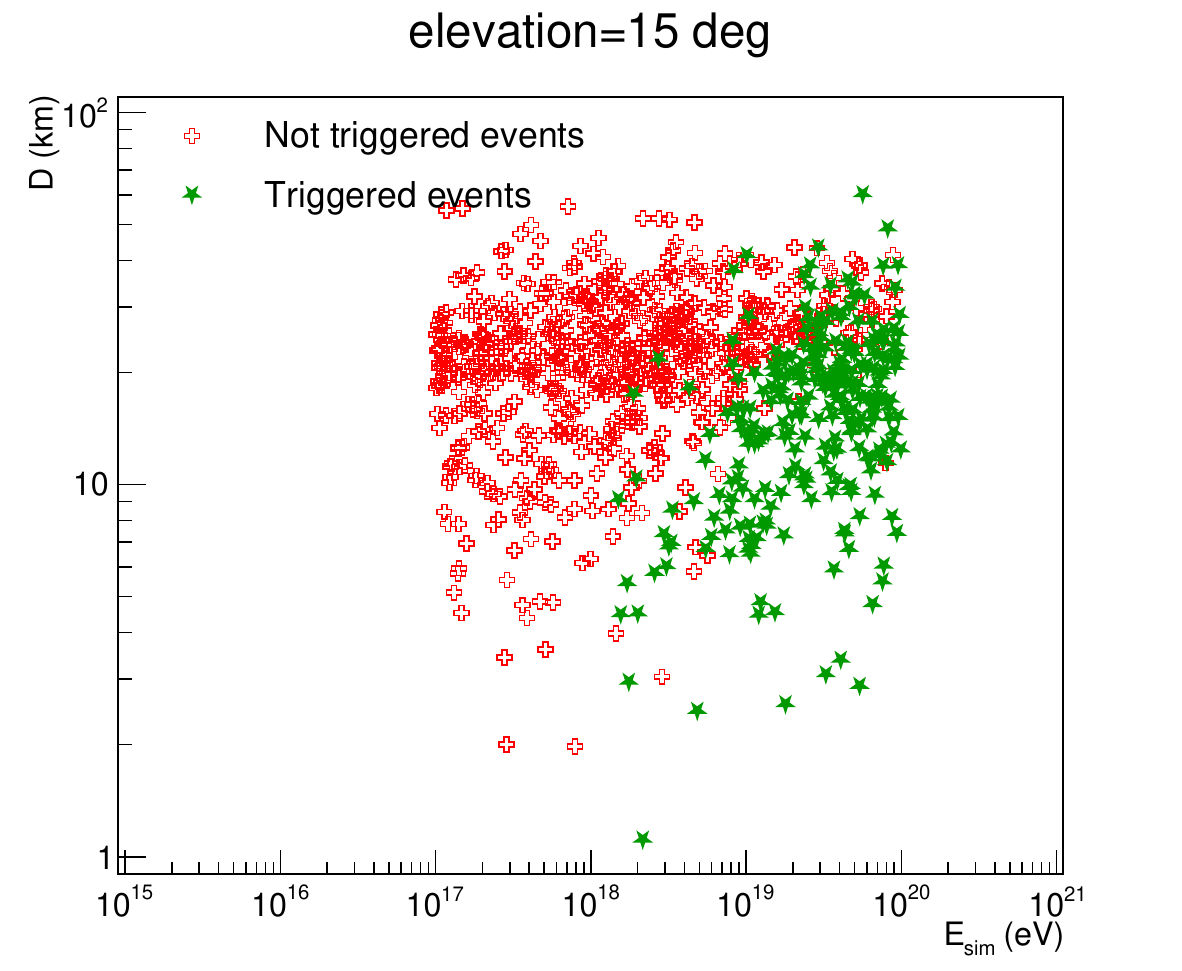}
	\includegraphics[width=6.cm]{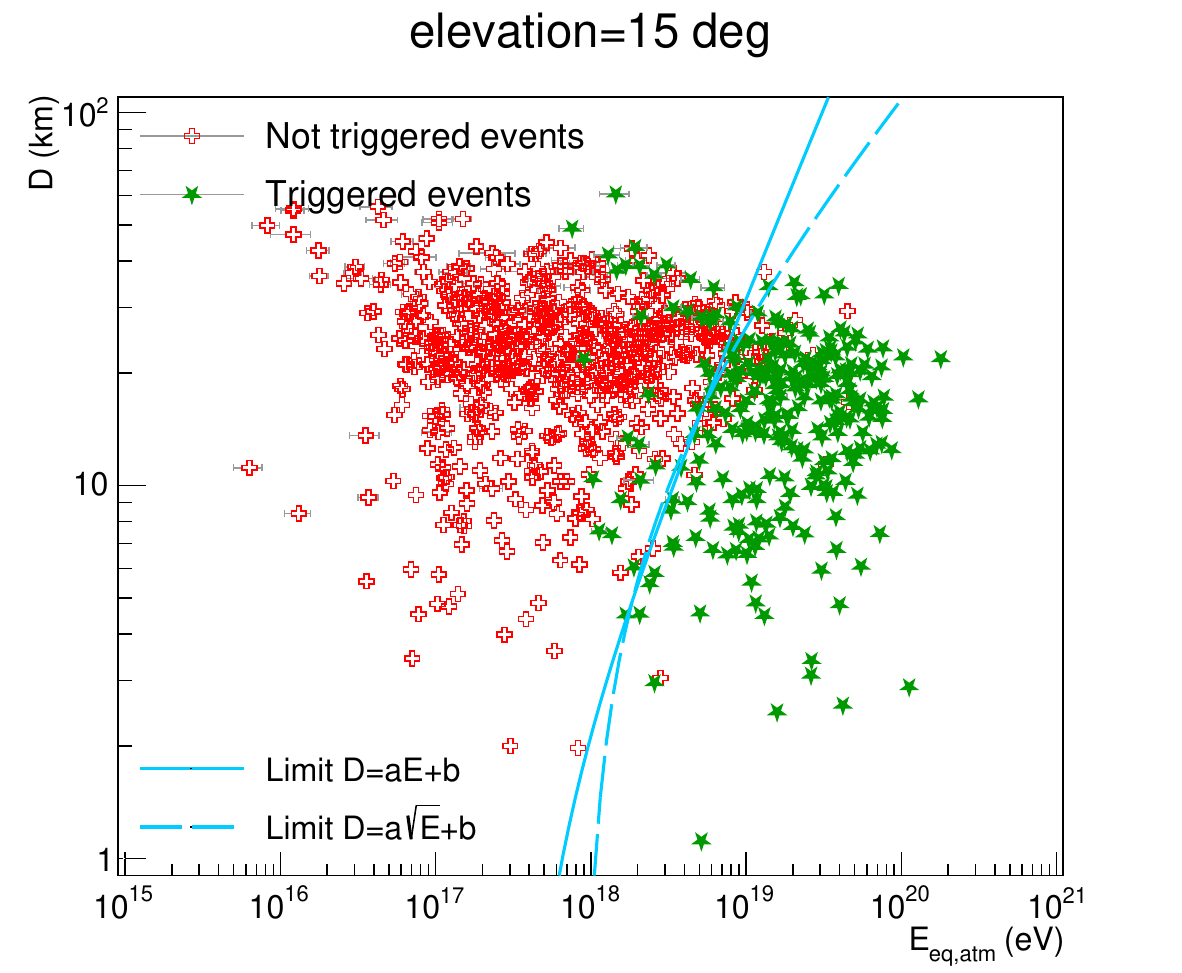}
	\caption{Triggered and non-triggered showers with their distance vs. the simulated energy (left) 
    and the equivalent energy corrected for the atmospheric transmission (right), for elevation angle \ang{15}, background 1~count/pixel/GTU, and SPACIROC3 board. The detection limits are drawn, too. } \label{fig:limit_Esim}
\end{figure*}

The simulation set described in Section~\ref{sec:simu_set} was used to study the detection limit.
In Figure~\ref{fig:limit_Esim} the simulated showers are plotted with their distance vs. simulated energy and equivalent energy corrected for the atmospheric transmission. 
Plots are for an elevation of $15^\circ$, background level of $1$~count/pixel/GTU, and SPACIROC3 electronics board. The same analysis was performed also for background level of $1.5$~count/pixel/GTU, elevation angles $10^\circ$, $20^\circ$, and $25^\circ$, and SPACIROC1.
The distance vs. $\mathrm{E_{sim}}$ plot (left) indicates that at greater distances, higher energies are needed to trigger the showers.
After the calculation of the equivalent energy corrected for the atmospheric transmission, in general, the energies decrease, as visible in the distance vs. $\mathrm{E_{eq,atm}}$ plot (right).

The detection limit should separate the triggered and the non-triggered showers that partially overlap, due to the variety of simulated showers, and to the fact that in the computation of the equivalent energy, only the fluorescence emission is taken into account. The Cherenkov emission, which is not considered in this analysis, can contribute to the signal and increase the detection probability.
The method defined to estimate the detection limit assumes that the triggered showers should lie on the right side of the detection limit (at higher energies), and the non-triggered ones on the left side. Therefore, an efficiency can be defined as $\mathrm{\epsilon=(N_{T}^{right}+N_{NT}^{left})/(N_{T}^{tot}+N_{NT}^{tot})}$,
where $\mathrm{N_{T}^{right}}$ and $\mathrm{N_{T}^{tot}}$ are the number of triggered showers on the right side of the line and the total number of triggered showers, respectively, and $\mathrm{N_{NT}^{left}}$ and $\mathrm{N_{NT}^{tot}}$ are the number of non-triggered showers on the left side of the line and the total number of non-triggered showers. The detection limit is defined by lines that maximize $\epsilon$.
As functions representing the limit, both the equations $\mathrm{D=aE+b}$ and $\mathrm{D=a\sqrt{E}+b}$ were considered, where $\mathrm{D}$ and $\mathrm{E}$ are the distance and the energy limit, respectively. 
The first equation takes into account that at greater distances, the signal integration time in a given pixel is greater. This reasoning leads to an energy limit $\mathrm{E\propto D}$, and vice-versa $\mathrm{D\propto E}$. 
The second equation is based on the fact that the registered counts decrease as $\mathrm{1/D^2}$ and that, therefore, the energy limit to register the minimum number of counts in order to trigger the shower is $\mathrm{E\propto D^2}$, and vice-versa $\mathrm{D\propto \sqrt{E}}$.
Iterations over different values of the angular coefficient $\mathrm{a}$ and of the intercept $\mathrm{b}$ led to the combination of parameters that maximize $\epsilon$ and defined the detection limit. Detection limits for these two distance-to-energy relationships are drawn in the plot on the right of Figure~\ref{fig:limit_Esim}. 

The detection limits were calculated for different backgrounds (1 and 1.5~counts/pixel/GTU), elevation angles ($10^\circ$, $15^\circ$, $20^\circ$ and $25^\circ$), and with SPACIROC1 and SPACIROC3 electronics boards. No clear improvement in the energy threshold was visible when using SPACIROC3 boards instead of SPACIROC1 boards, but all the results together define a range of energy varying with the distance.

\section{Estimation of the level-1 trigger rates}\label{sec:trigger_sim}

The foreseen upgrade of EUSO-TA includes the self-trigger capability, to operate independently from the TA-BRM-FDs. The level-1 trigger logic for the observation of showers, was designed and implemented in the \Offline framework, in order to test it through simulations and evaluate the UHECRs trigger rate. 
The simulation sets discussed in Section~\ref{sec:simu_set} were used also in this analysis. 

\begin{figure*}[h!]
	\centering
	\includegraphics[width=5.cm]{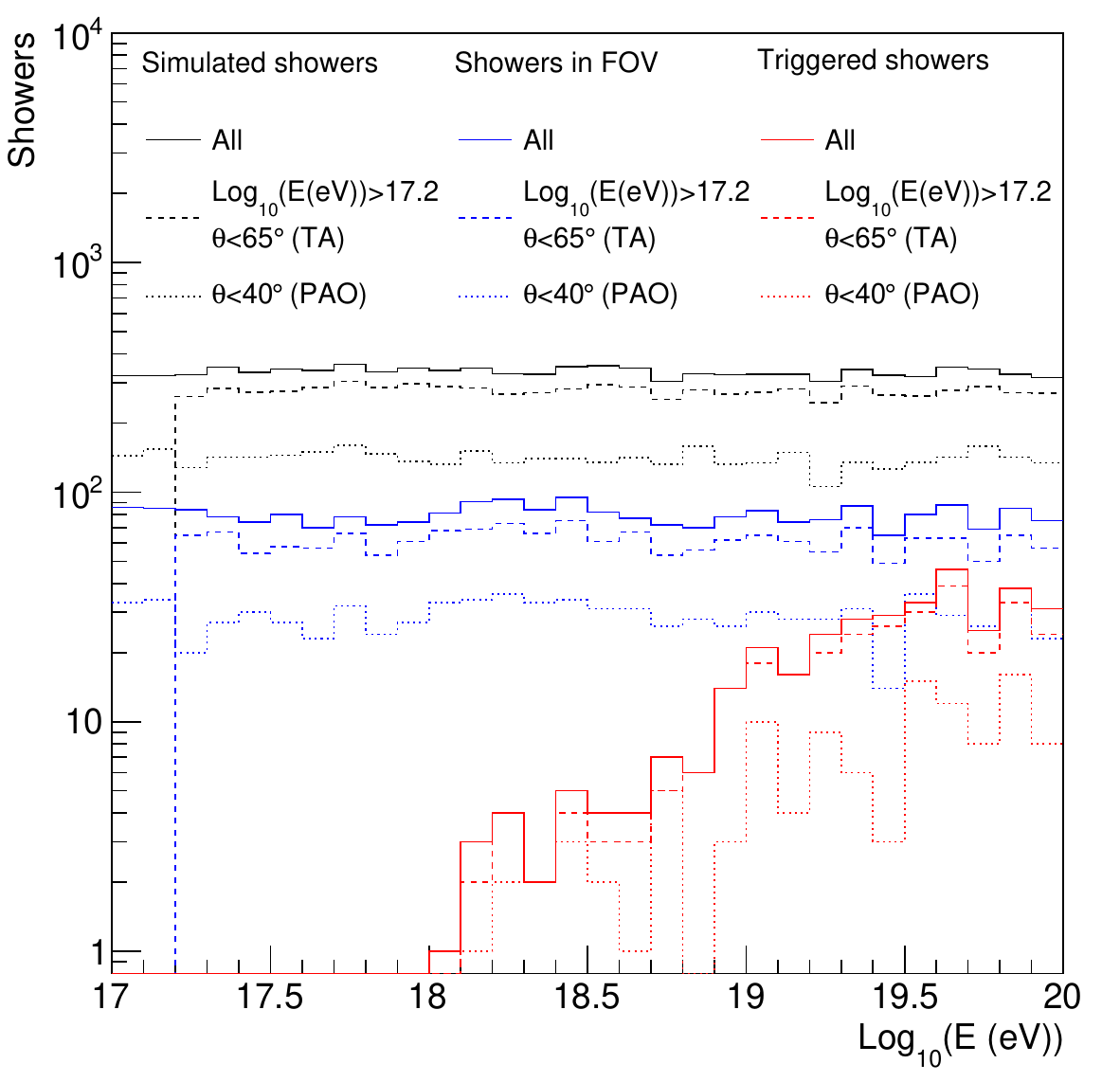}
	\includegraphics[width=5.cm]{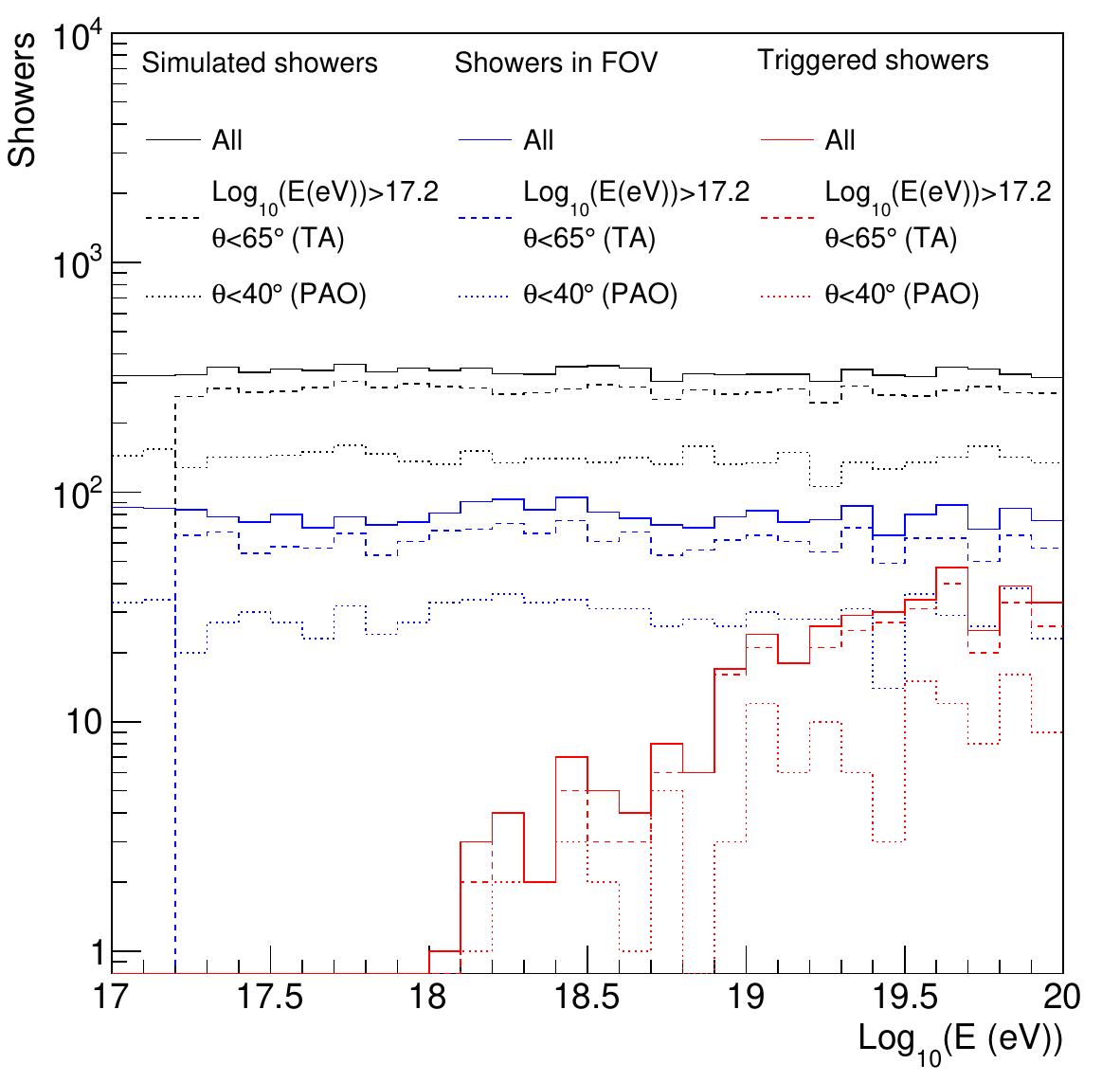}\\
	\includegraphics[width=5.cm]{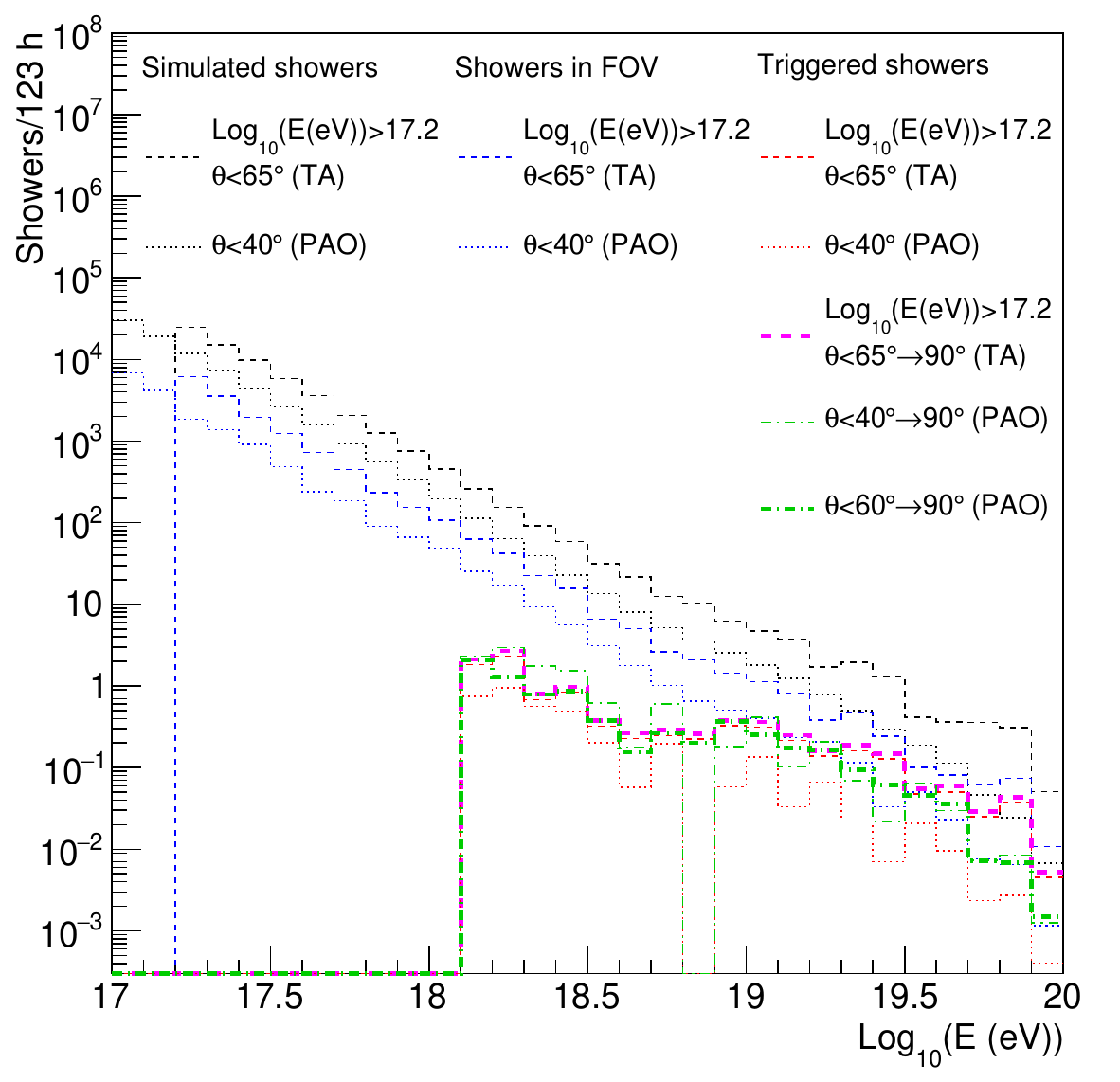}
	\includegraphics[width=5.cm]{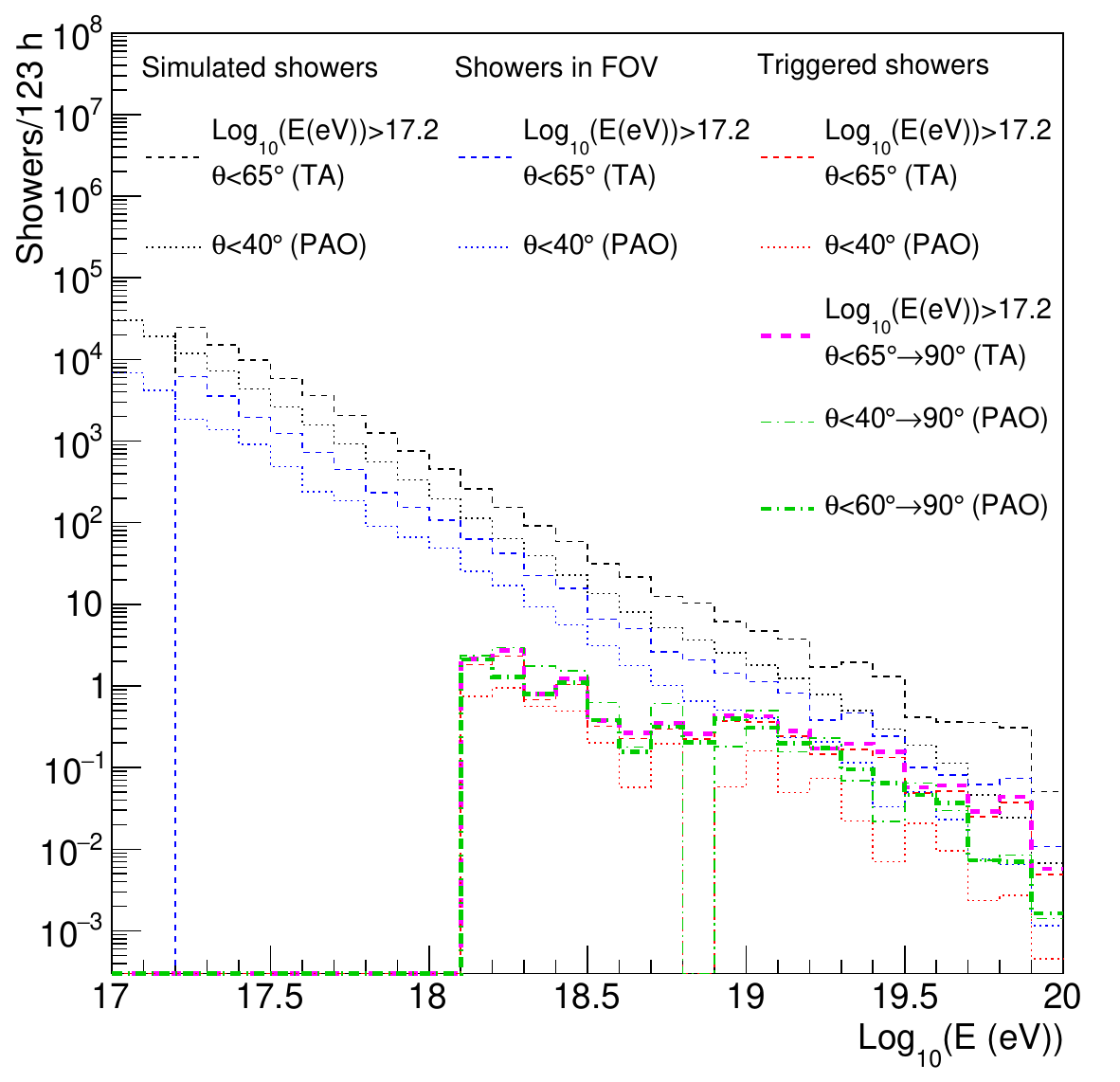}
	\caption{Energy distribution of the simulated showers with spectral index $-1$ in logarithmic scale (top), and rescaled by the UHECR flux measured by the PAO and the TA (bottom) for elevation angle $15^\circ$ and for SPACIROC1 (left) and SPACIROC3 (right) boards. The distributions are for the simulated showers, showers in the FOV, and triggered showers. Solid lines are for all showers with no cuts; dashed lines show the distributions with the cuts on the energy and zenith angle used to measure the flux by TA; dotted lines show the distributions with the cuts on the zenith angle used to measure the flux by PAO.} \label{fig:offline_sim_energy_histo}
\end{figure*}

In Figure~\ref{fig:offline_sim_energy_histo}, the energy distributions of the simulated showers with spectral index $-1$ in logarithmic scale are shown (top). The plots are for elevation angle $15^\circ$ and for both the SPACIROC1 (left) and SPACIROC3 (right). The distributions represent the simulated showers (black), the showers in the FOV (blue), and the triggered showers (red). 
The energy distributions must be rescaled based on measured fluxes of UHECRs, to retrieve the natural spectral index $-2$ in logarithmic scale. For this purpose, both the energy fluxes measured with the Telescope Array (TA) \cite{bib:fluxTA} and the Pierre Auger Observatory (PAO) \cite{bib:fluxPAO_40} were used. The flux measured by TA included showers with energy $\mathrm{E>10^{17.2}}$~eV and zenith angles $\theta<65^{\circ}$, while that measured by the PAO included showers with zenith angles $\theta<40^{\circ}$.
In the plots, solid lines are for all showers with no cuts; dashed lines show the distributions with the cuts used to measure the TA flux; the dotted lines show those with the cuts used for the PAO flux.
It is visible that showers start to generate triggers at energies $\simeq10^{18}$~eV.

In the first step, the energy distributions with the same cuts used in the TA and PAO spectra were rescaled. 
For each $i$-th bin, the number of expected triggered events $\mathrm{n_{trig,exp,i}^{cut}}$ with cuts and in the time interval of 123~h (observation time during the data acquisitions in 2015, used as a reference), is calculated as $\mathrm{n_{trig,exp,i}^{cut}=n_{trig,i}^{cut}/n_{sim,i}^{cut}\cdot S_{EUSO-TA}^{cut} \cdot F_{i}^{cut}\cdot dE_i}$
where $\mathrm{n_{trig,i}^{cut}}$ and $\mathrm{n_{sim,i}^{cut}}$ are the number of triggered and simulated events with cuts, respectively; $\mathrm{S_{EUSO-TA}^{cut}}$ is the exposure of EUSO-TA; $\mathrm{F_{i}^{cut}}$ is the flux measured with the PAO or the TA; $\mathrm{dE_i}$ is the differential energy. The rescaled distributions are visible in the bottom plots of Figure~\ref{fig:offline_sim_energy_histo}.
%
%

In the second step, the expected trigger rates with cuts were rescaled to retrieve those without cuts on the zenith angle $\mathrm{n_{trig,exp,i}}$. This was done by multiplying the first with the ratio between the total number of events $\mathrm{N_{trig}}$ and the total number of events with cuts $\mathrm{N_{trig}^{cut}}$:  $\mathrm{n_{trig,exp,i}=N_{trig}/N_{trig}^{cut}\cdot n_{trig,exp,i}^{cut}.}$
The corresponding distributions are plotted in the bottom plots of Figure~\ref{fig:offline_sim_energy_histo} with magenta and green dashed lines, considering the TA and PAO spectrum, respectively. The cut on the zenith angle used for the PAO spectrum ($\theta <40^\circ$) in some cases was considerably reducing the statistics of events, as can be seen in the distribution of triggered events with this cut in the bottom plots of Figure~\ref{fig:offline_sim_energy_histo}, where one bin is empty. As the flux per bin measured by the PAO with $\theta <40^\circ$ differs only by a few percent from that measured with the same experiment with $\theta <60^\circ$ \cite{bib:fluxPAO_60}, the flux with $\theta <40^\circ$ was applied to the larger set of events with $\theta <60^\circ$\footnote{The flux in \cite{bib:fluxPAO_60} was limited to energies $\mathrm{E>10^{18.4}}$~eV, and to simplify the procedure and avoid problems of normalization between two sets of low- and high-energy spectra, the flux in \cite{bib:fluxPAO_40} was considered in this study.}. The final distributions of triggered events rescaled with the TA and PAO fluxes are presented in thicker lines, in magenta and green, respectively. 
The potential trigger rate in 123~hours was calculated as the sum of the expected events per bin.

Thinking in terms of single acquisition sessions (about 30~h each), these results indicate that with the level-1 trigger it could be possible to detect a few~showers/session in case of 1~count/pixel/GTU background level and SPACIROC1 boards. Using SPACIROC3 boards, the trigger rates increase by 15--20\%. With a higher background of 1.5~counts/pixel/GTU, the trigger rates halves. 
For reference, four showers were found by applying (offline) the level-1 trigger algorithm on the data collected with the external trigger in 2015, which corresponds to about $1$~shower/session. Including showers recognized with manual checks, a total of nine showers were identified, or about 2~showers/session.

\section{Summary and Conclusion}\label{sec:summary}
In order to evaluate the detection limit of EUSO-TA and the expected trigger rate with the level-1 trigger developed for EUSO-TA, several configurations have been considered, in terms of electronics (SPACIROC1 boards that were used during the data taking in 2015 and 2016 and SPACIROC3 included in the upgrade of the detector), and two background levels (1 and 1.5~counts/pixel/GTU).

To evaluate the detection limit, the real energy of showers have been rescaled to account for the fact that only a portion of them can be observed due to its limited FOV (with respect to usual cosmic ray fluorescence detectors on ground). This rescaling was applied to the different sets of simulations. For all configurations, the detection limit was evaluated and in general ranges between  $1\times10^{18}-1\times10^{19}$~eV in between 0 and 50~km.

The expected trigger rate for EUSO-TA with the internal level-1 trigger was evaluated too, in the same simulated conditions. The expected rate is a few detections per acquisition session (of about 30~h each), in the case of low background level (1~count/pixel/GTU) and SPACIRCOC1 boards. By using SPACIROC3 boards, the trigger rates increase by $15\%-20\%$. For the higher background level (1.5~counts/pixel/GTU) the rates halve.

A more complete description of the analysis and more detailed results will be included in a paper under preparation.
\\

{
\noindent\small{\bf Acknowledgments:}
This work was partially supported by Basic Science Interdisciplinary Research
Projects of RIKEN and JSPS KAKENHI Grant (JP17H02905, JP16H02426 and
JP16H16737), by the Italian Ministry of Foreign Affairs and International Cooperation,
by the Italian Space Agency through the ASI INFN agreements Mini-EUSO n. 2016-1-U.0, EUSO-SPB1 n. 2017-8-H.0, OBP (n. 2020-26-Hh.0), EUSO-SPB2 n. 2021-8-HH.0 and by ASI INAF agreeement n. 2017-14-H.O, by NASA awards and grants 11-APRA-0058, 16-APROBES16-0023, 17-APRA17-0066, NNX17AJ82G, NNX13AH54G, 80NSSC18K0246, 80NSSC18K0473, 80NSSC19K0626, 80\-NSSC\-18K\-0464 and 80NSSC22K1488 in the USA, Deutsches Zentrum f\"ur Luft- und Raumfahrt, by the French space agency
CNES, the Helmholtz Alliance for Astroparticle Physics funded by the
Initiative and Networking Fund of the Helmholtz Association (Germany), by National Science Centre in Poland grant no 2017/27/B/ST9/02162 and
2020/37/B/ST9/01821. L. W. Piotrowski acknowledges financing by the Polish National Agency for Academic Exchange within Polish Returns Programme no. PPN/PPO/2020/1/00024/U/00001 and National Science Centre, Poland grant no. 2022/45/B/ST2/02889.
Russian team is supported by ROSCOSMOS, "KLYPVE" is included into the
Long-term program of Experiments on board the Russian Segment of the ISS.
Sweden is funded by the Olle Engkvist Byggm\"astare Foundation.
}

%

\newpage
{\Large\bf Full Authors list: The JEM-EUSO Collaboration\\}

\begin{sloppypar}
{\small \noindent
S.~Abe$^{ff}$, 
J.H.~Adams Jr.$^{ld}$, 
D.~Allard$^{cb}$,
P.~Alldredge$^{ld}$,
R.~Aloisio$^{ep}$,
L.~Anchordoqui$^{le}$,
A.~Anzalone$^{ed,eh}$, 
E.~Arnone$^{ek,el}$,
M.~Bagheri$^{lh}$,
B.~Baret$^{cb}$
% ...
}
\end{sloppypar}
\vspace*{.3cm}

{ \footnotesize
\noindent
% ...
$^{cb}$ Universit\'e de Paris, CNRS, AstroParticule et Cosmologie, F-75013 Paris, France\\
% ...
$^{ed}$ Istituto Nazionale di Fisica Nucleare - Sezione di Catania, Italy\\
% ...
$^{eh}$ INAF - Istituto di Astrofisica Spaziale e Fisica Cosmica di Palermo, Italy\\
% ...
$^{ek}$ Istituto Nazionale di Fisica Nucleare - Sezione di Torino, Italy\\
$^{el}$ Dipartimento di Fisica, Universit\`a di Torino, Italy\\
% ...
$^{ep}$ Gran Sasso Science Institute, L'Aquila, Italy\\
% ...
$^{ff}$ Nihon University Chiyoda, Tokyo, Japan\\ 
% ...
$^{ld}$ University of Alabama in Huntsville, Huntsville, AL, USA\\
$^{le}$ Lehman College, City University of New York (CUNY), NY, USA\\
% ...
$^{lh}$ Georgia Institute of Technology, USA\\
% ...
}

%
%
%

\end{document}